# Classify Sina Weibo users into High or Low happiness Groups Using Linguistic and Behavior Features

Jingying Wang[1], Tianli Liu[2], Tingshao Zhu[1,*], Lei Zhang[3], Bibo Hao[1], and Zhenxiang Chen[3]

1 {Institute of Psychology, University of Chinese Academy of Sciences}, CAS, Beijing, China
2 Institute of Population Research, Peking University, Beijing, China
3 Shandong Provincial Key Laboratory of Network Based Intelligent Computing,
University of Jinan, Jinan, China

**Abstract**: It's of great importance to measure happiness of social network users, but the existing method based on questionnaires suffers from high costs and low efficiency. This paper aims at identifying social network users' happiness level based on their Web behavior. We recruited 548 participants to fill in the Oxford Happiness Inventory (OHI) and divided them into two groups with high/low OHI score. We downloaded each Weibo user's data by calling API, and extracted 103 linguistic and behavior features. 24 features are identified with significant difference between high and low happiness groups. We trained a Decision Tree on these 24 features to make the prediction of high/low happiness group. The decision tree can be used to identify happiness level of any new social network user based on linguistic and behavior features. The Decision Tree can achieve 67.7% on precision. Although the capability of our Decision Tree is not ideal, classifying happiness via linguistic and behavior features on the Internet is proved to be feasible.

# Introduction

Positive psychology is a subject which focuses on positive traits of human beings (Seligman & Csikszentmihalyi, 2000). Since both distress and happiness co-exist in real world, positive psychologists argued that distress is not the only thing that psychology cares, happiness should be paid attentions as well (Peterson, 2006). The true value of positive psychology is to study normal person and assist healthy people to achieve their value of life. Happiness plays a key role in positive psychology since people care about it (Peterson, 2006). In order to assess happiness, many measurement methods have already been developed. Some questionnaires are widely used, including The Oxford Happiness Inventory (OHI, Argyle), The Satisfaction With Life Scale (SWLS, Diener), The Scale of Psychological well-being (SPWB, Ryff), The Positive and Negative Affect Schedule (PANAS, watson), etc. There are two mostly used methods, self-report and experience sampling method (ESM). The self-report is a traditional method for assessment in psychological studies, although it is reported to work with high reliability and validity, it is weak to deal with large sample, and susceptible to the deviation of memory at the same time. ESM can receive real-time respond, thus avoid the deviation of memory, but it is very expensive to carry out large-scale studies. In recent years, social media has grown rapidly, attracting more and more people to express themselves freely, which makes it an ideal platform to collect mass data in short time with low costs, and it also attracted more and more research conducted on social media.

---

* Corresponding Author: Email: tszhu@psych.ac.cn

It is reported there exists significant correlation between online behaviors and well-being. Kraut. R (1998) found that using Internet just three hours per week would lead to high depression and low well-being. Weiser (2001) reported that different way of Internet use could cause different effects on human well-being. Online self-disclosure is found to have positive impact on well-being when people obtained a better interpersonal relationship via online self-disclosure (Valkenburg & Peter, 2009). The result of Caplan' studies (2007; 2009) showed that online games have a positive relationship with Problematic Internet Use (PIU), and PIU have a remarkable influence on depression, anxiety and loneliness, which reduce well-being in consequence.

Recently, there are several studies found some interesting behavioral patterns on social media according to well-being, such as characterizing geographic variation in well-being (Hansen Andrew Schwartz et al., 2013), linguistic differences of people online with different personalities, gender and age (H. Andrew Schwartz et al., 2013) and so on. Prediction of psychological traits via SNS became popular, like personality (Adalı & Golbeck, 2014; Li et al., 2014), suicide risk (Poulin et al., 2014; Zhang et al., 2014), well-being (Kross et al., 2013) etc.

In this paper, we propose to investigate how people with different happiness level behave on social media, and some preliminary results will be presented and discussed. It is expected that these linguistic and behavioral differences may lead to more efficient method to identify happiness on social media.

# Method

The primary goal of this research was to identify the features which could distinguish different levels of happiness in microblog users. The final goal was to utilize these features to predict users' happiness. In this section, we describe how we selected participants, carried out the study, and collected the data.

## Participants

We recruited Sina Weibo users meeting the following criteria: (1) they must be active users who have posted at least 500 tweets in their status updates before our recruitment; (2) they have to complete two online questionnaires including a demographic scale and a scale of happiness; (3) by accepting a consent, they agree to provide us with access to their microblogs and use their data anonymously in an automated setting, without active human intervention, in doing research analyses. A total of 2,004 users completed our investigations from September 26, 2012 to October 27, 2012. And then we excluded respondents whose questionnaires were invalid or who refused to accept the consent, which yielded a sample of 1773 users. In order to fulfil our research purpose of characterizing happiness, we selected subjects whose score of happiness either located one standard deviation (SD) above the mean (M) or one SD below the M in the sample distribution. Finally, 548 subjects were included in and they were divided into two groups. Two groups of participants were constructed in this manner: the HH group of 294 users scoring high (M+SD) for happiness; and LH group of 254 users who have low scores (M - SD) of happiness. The sample contained more females (345, 63%) than males (203, 37%), with a mean age of 23.8.

## Procedure

Sina Weibo is one of the most popular social network sites in China, similar to Twitter. Using our public account, we randomly send inviting messages to 20,000 active users to participate the study. For the purpose of obtaining enough information to analyze, we defined active users as those who have posted at least 500 tweets in their accounts before recruitment. Every participant visited the study website

(http://ccpl.psych.ac.cn:10002/) via a link we written on the inviting message. They used their Weibo accounts to complete a registration. When participants logged into the study website, they completed an online consent form. After agreeing to participate in the study, they were asked to provide demographic information such as gender, age etc. the main survey they need to fill in was Oxford Happiness Inventory (OHI). After one month of recruitment, we downloaded each Weibo user's data by calling Sina Weibo API.

Before the variation analysis, knowing the types of the data distribution is necessary for us to decide which statistical methods should be used in further analysis. we computed the mean (M=87.6) and SD (=21.6) of the scores of OHI, then classified subjects as HH group (> M+SD=109.2) or LH group (< M+SD=66.0) based on their scores. The next, normality test was used to determine whether every feature follow a normal distribution. The results of test showed that none of these variables obey normality. Based on the results of the normality test and the rules of psychometrics, we chose Mann–Whitney U test to analyze the differences between the two groups.

We divided our study into two steps for achieving the goal of classification. We first analyzed the differences in demographics, linguistic and behavior features between LH group and HH group. Linguistic features were obtained by Simplified Chinese version LIWC (SCLIWC, 2013) analysis of the users' social media postings before the onset of recruitment. Behavioral features, the features relate to users' behaviors of Weibo using, most of them can be directly collected from the profiles of users' accounts. The other two features were calculated by computer. Then statistical tests were employed to examine the differences between the LH and HH groups. Subsequently, we leveraged these linguistic and behavior features, to build a statistical classifier that provides estimates of the level of happiness. The linguistic and behavior features were used to build predictive models through SPSS 22.0.

## Data collection

In this section, we discuss the data we needed and the process we collected data. The data of participants directly downloaded from Sina Weibo API after the survey one month later, which covers microblogs from the first Weibo to the last Weibo which posted one month after our finished the recruitment. The dataset comprised linguistic and behavior features which extracted from the data of Weibo.

**Measures**

Demographic data and a measure of happiness were collected from the participants on Internet. Demographics including age, gender, degree of education, marital status, place of residence, income, health state, religion, eight variables in total. The Oxford Happiness Inventory (OHI) (Argyle, Martin, & Crossland, 1989) was used to measure happiness, and the results of OHI were regarded as the gold standard of this study. OHI has 29 items, each item contains four gradually upgrade emotional sentences. Total point equal to sum of each item's score. Higher scores represent higher happiness. The Chinese version of OHI has been widely applied in psychological studies in China and has shown very good reliability and validity (Shi, Wang, & Deng, 2005; Wang & Wang, 2011).

**Linguistic features**

In working toward finding linguistic features characterizing happiness, we focus on lexical features which are easily interpreted by SCLIWC. SCLIWC was a modified Chinese version of LIWC (linguistic inquiry and word count, Pennebaker, et al., 2007)[1], which is a common text analysis software program designed for linking language with psychological variables involves counting words belonging to categories of language. LIWC includes four main types of linguistic features, including process of language, mental process, personal concern and oral language, refer to negative emotion, positive emotion, family, work, etc. for a total of 88 categories. *LIWC* counts how often words in a given category are used by an individual, and calculates the percentage of the words which are from the given category. Based on the cyber language in Sina Weibo, SCLIWC expanded 4560 new words which have high-frequency usage but outside the scope of the lexicon of LIWC, and next the words were judged whether it could left or not by psychological experts.

---
[1] http://www.liwc.net/

**Behavior features**

Human behavior is a mirror reflects a psychological world. Behaviors can also reveal the relationships among people, especially on social network sites. For instance, two users mutually follow each other, they are friends in reality. Take account of the practical needs and privacy protection, we defined three kinds of behaviors:

- *About personal private*: whether anyone was allowed to send private messages to me, whether anyone was allowed to comment my microblogs, has a custom avatar, turn on/off the geo-location and whether the account was verified.
- *About online social contact*: the amount of user' following, the amount of user' followers, mutual followers count, mutual followers count divide followers count and mutual followers count divide following count.
- *About self-expression*: the length of self-description, whether the self-description contains the word "me", favorites count, statuses count and the ratio of original microblogs.

The main categories of behavior features that we consider in this paper are described below (see table 1):

**Table 1**. Behavior features and their definitions

| Feature | Definition |
| --- | --- |
| Allow all messages | If anyone allowed to send private messages to me |
| Allow all comments | If anyone allowed to comment my microblogs |
| Avatar | Has a custom avatar |
| Mut-followers /followers | Mutual followers count/followers count |
| Mut-followers /following | Mutual followers count/following count |
| Mut-followers count | Mutual followers count |
| Description | Length of Self-description |
| Description me | Self-description including the word "me" |
| Favorites count | Favorites count |
| Followers count | Followers count |
| Following count | Following count |
| Geo-enabled | Turn on/off geo-location |
| Statuses count | Statuses count |
| verified | If verified or not |
| Ratio of original Weibo | The ratio of original microblogs |

# Results

## Sample characteristics

The education attainment of the study sample (83.2% had completed at least some college, and 29% had a bachelor's degree or more) is higher than that of general population (the largest proportion of education degree is "middle school", accounts for 38.8%, based on the sixth China population census in 2010). Income was unevenly distributed across six income blocks, which ranged from "under ￥2000 per month" to "over ￥10,000 per month", two thousands RMB each block, and the most frequently reported income range was "￥2,000-￥4,000", accounted for 35%. In marital status, 65.3% were unmarried. 58.6% of the participants lived in municipalities directly under the central government or provincial capitals, 35.8% lived in common cities, only 5.7% lived in countries. Most participants (83.6%) were healthy. 83.9% of them were not believers of religion.

## Features related to happiness

In the light of the above descriptions, we present an exploration of linguistic and behavior features of high happiness and low happiness classes.

Table 2 lists demographic characteristics of the HH group and the LH group. We found that the HH group were more likely to be married and healthy, and to report higher education, high-income and religious beliefs more than the LH group. In the area of degree of education, 11.5% had completed at most specialized secondary school in the HH group, comparing 22.8% in the LH group. Income distribution of the LH group displayed that the first two blocks accounted for a combined six out of ten (22.4% + 38.2%). The other side, people whose income below 4000 RMB only possessed 42.7% (32.3% + 10.5%) in the HH group, far less than the LH group. There was also a significant contrast on "place of residence" between the two groups: a larger proportion (65%) of people live in municipalities directly under the central government or provincial capitals in the HH group, compared to 51.2% in the LH group; only 2.3% live in villages and towns in the HH group, in contrasted with 9.4% in the LH group.

Table 2. Differences of demographic features

| Variable | HH group | | LH group | | $\chi^2$ | df | P |
|---|---|---|---|---|---|---|---|
| | N | % | N | % | | | |
| **Degree of education** | | | | | 230.63 | 7 | .000 |
| Elementary or below | 0 | 0 | 0 | 0 | | | |
| Junior school | 1 | 0.3 | 13 | 5.1 | | | |
| Senior school | 26 | 8.8 | 27 | 10.6 | | | |
| Specialized secondary school | 7 | 2.4 | 18 | 7.1 | | | |
| Junior college | 181 | 61.6 | 116 | 45.7 | | | |
| Bachelor | 48 | 16.3 | 70 | 27.6 | | | |
| Master | 26 | 8.8 | 10 | 3.9 | | | |
| Doctor | 5 | 1.7 | 0 | 0 | | | |
| **Marital status** | | | | | 14.69 | 1 | .000 |
| Singlehood | 176 | 59.9 | 182 | 71.7 | | | |
| Married | 118 | 40.1 | 72 | 28.3 | | | |
| **Place of residence** | | | | | 67.57 | 3 | .000 |

| | | | | | | | | |
|---|---|---|---|---|---|---|---|---|
| Village | | 1 | 0.3 | 5 | 2.0 | | | |
| Town | | 6 | 2.0 | 19 | 7.5 | | | |
| Common city | | 96 | 32.7 | 100 | 39.4 | | | |
| Municipality | | 191 | 65.0 | 130 | 51.2 | | | |
| **Income** | | | | | | 54.10 | 5 | .000 |
| Below 2,000¥ | | 31 | 10.5 | 57 | 22.4 | | | |
| 2,000-4,000 | | 95 | 32.3 | 97 | 38.2 | | | |
| 4,000-6,000 | | 65 | 22.1 | 37 | 14.6 | | | |
| 6,000-8,000 | | 35 | 11.9 | 28 | 11.0 | | | |
| 8,000-10,000 | | 24 | 8.2 | 16 | 6.3 | | | |
| Above 10,000 | | 44 | 15.0 | 19 | 7.5 | | | |
| **Health state** | | | | | | 459.99 | 1 | .000 |
| Unhealthy | | 12 | 4.1 | 78 | 30.7 | | | |
| Healthy | | 282 | 95.9 | 176 | 69.3 | | | |
| **Religion** | | | | | | 10.05 | 1 | .002 |
| Non-believer | | 236 | 80.3 | 224 | 88.2 | | | |
| Believer | | 58 | 19.7 | 30 | 11.8 | | | |

As shown in Table 3, we found the HH group used some categories of words more frequently than the LH group, such as person plurals, "work", "religion" etc.

**Table 3.   Differences of linguistic features**

| Categories | HH group | | LH group | | Mann-Whitney U | Wilcoxon W | Z | P |
|---|---|---|---|---|---|---|---|---|
| | M | SD | M | SD | | | | |
| First-person plural | 0.14 | 0.28 | 0.12 | 0.28 | 33200.500 | 65585.500 | -2.409 | .016 |
| Second-person plural | 0.05 | 0.14 | 0.03 | 0.09 | 32361.500 | 64746.500 | -3.560 | .000 |
| Preposition | 3.62 | 1.48 | 3.37 | 1.64 | 33589.500 | 65974.500 | -2.028 | .043 |
| Third-person plural | 0.14 | 0.35 | 0.07 | 0.16 | 31028.000 | 63413.000 | -3.842 | .000 |
| Multifunction | 6.43 | 2.23 | 5.95 | 2.81 | 32283.000 | 64668.000 | -2.735 | .006 |
| Social processes | 5.62 | 2.37 | 5.16 | 2.55 | 32713.500 | 65098.500 | -2.502 | .012 |
| Humans | 1.35 | .89 | 1.27 | 1.20 | 33556.000 | 65941.000 | -2.047 | .041 |
| Certain | 1.49 | .86 | 1.38 | 1.00 | 33478.500 | 65863.500 | -2.088 | .037 |
| Space | 3.95 | 1.63 | 3.61 | 1.67 | 32049.000 | 64434.000 | -2.862 | .004 |
| Work | 1.74 | 1.61 | 1.38 | 1.41 | 30905.000 | 63290.000 | -3.482 | .000 |
| Achieve | 0.99 | 0.82 | 0.88 | 0.83 | 33611.500 | 65996.500 | -2.019 | .043 |
| Religion | 0.51 | 0.55 | 0.40 | 0.43 | 33011.500 | 33000.500 | -2.358 | .018 |
| Nonfluency | 0.91 | 0.73 | 0.83 | 1.01 | 65396.500 | 65385.500 | -2.352 | .019 |

We observed that the HH group have more counts of "follower", "following", "mutual followers" and a higher ratio of mutual followers and following. The feature "verified" means whether the account was obtained a certified identity. The percentage of "verified" in the HH group was 4.4% (thirteen users), while the LH group was close to zero (0.4%, only one user had verified), which suggests more people in the HH group were certified their real identity.

Table 4. Differences of behavior features

| Type of behavior | HH group | | LH group | | Mann-Whitney U | Wilcoxon W | Z | P |
|---|---|---|---|---|---|---|---|---|
| | M | SD | M | SD | | | | |
| Mut-followers /following | 0.43 | 0.24 | 0.38 | 0.23 | 32990.00 | 65375.00 | -2.35 | .019 |
| Mut-followers count | 156.26 | 183.99 | 111.78 | 134.38 | 29947.50 | 62332.50 | -3.99 | .000 |
| Followers count | 870.16 | 2001.73 | 679.13 | 1715.16 | 30324.50 | 62709.50 | -3.79 | .000 |
| Following count | 373.23 | 322.35 | 317.81 | 302.05 | 32147.50 | 64532.50 | -2.81 | .005 |
| Verified | 0.04 | 0.21 | 0.00 | 0.06 | 35834.00 | 68219.00 | -2.98 | .003 |

## Classification of happiness

We first presented a set of demographic characteristics as well as linguistic and behavioral features associated with level of happiness. Next we utilized characteristics and features to predict their levels of happiness. For all of our analyses, we used both 10-fold cross validation on the set of 548 participants, and compared several different feature sets to determine the best classification effects. Considered the LH group as the positive example, the results indicated that the best performing model in our test set yields an average accuracy of ~60% and precision of 0.677 (see Table 5). In order to understand the importance of various feature types, we trained a few models. The results with which we can predict happiness using different feature sets are present in Table 5. As shown below, the behaviors significantly improve prediction precision. The behavior feature set is more predictive than the linguistic feature set, almost equal to the effect of the linguistic and behavior feature set. The linguistic features do not seem to add additional predictive value when used in addition to the behavior feature set. All three feature sets combined give more accurate results than the other models, which yielded the value of accuracy arrived at ~65%.

Table 5. Results of Classification via Decision Tree

| | Precision | | Recall | | F-Measure | | Accuracy | |
|---|---|---|---|---|---|---|---|---|
| Software | SPSS | Weka | SPSS | Weka | SPSS | Weka | SPSS | Weka |
| L | 0.157 | 0.490 | 0.104 | 0.508 | 0.125 | 0.499 | 0.578 | 0.527 |
| B | 0.795 | 0.520 | 0.699 | 0.831 | 0.744 | 0.639 | 0.575 | 0.566 |
| L+B | 0.677 | 0.527 | 0.553 | 0.500 | 0.609 | 0.513 | 0.611 | 0.560 |
| L+B+D | 0.307 | 0.632 | 0.164 | 0.650 | 0.214 | 0.641 | 0.657 | 0.662 |

Notes: L=linguistic features, B=behavior features, D=demographic features. Test set prediction happiness using L, B, D, and all three together, as measured using Decision Tree via SPSS and Weka

For the purpose of understanding these features contributing to happiness, we draw a path diagram via Decision Tree. In figure 1, the pathway of tree graph displays that mutual follower count, second-personal plural pronouns and category of "humans" are the top three effective factors.

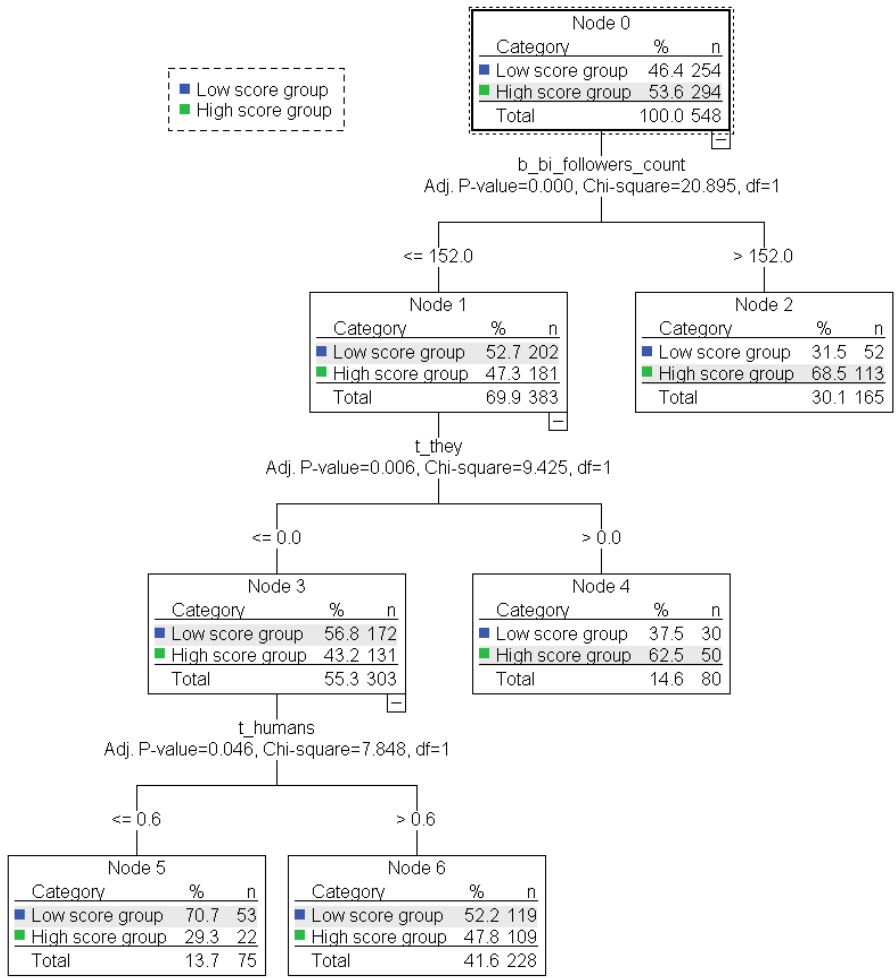

**Figure 1**. A pathway of low-happiness: four levels

# Discussion

In this section, we discuss two questions: first of all, what caused these distinctions between the two groups? secondly, which features have more contribution to the predictive proficiency in our model?

**Differences.** The findings of this study revealed a demographics pattern for happiness that higher level of happiness is associate with being richer, married, healthy, religious believer and living in metropolis. In light of *the Easterlin Paradox* and a comparative study of happiness across different nations, increased income would markedly improve the living level of citizens in an underdeveloped area (Diener & Oishi, 2000; Easterlin, 1974). In this study, the holistic income of the HH group is higher than the LH group. China is a developing country with a low per capita income, that is why improved earning still plays a role in promoting happiness. We infer that the condition of residence has strong impact on one's quality of life. Living in municipalities or provincial capitals is conducive to finding a good job, accepting better education and obtaining more social resources. These benefits could be sources of happiness, which was confirmed by the previous study (Adams, 1992).

As external causes, major life events such as marital status, health state and income (financial situation) all connect with happiness (Headey & Wearing, 1989). The differences of marital status between the two groups may be explained by that marriage as an important life event has a positive correlation with well-being (Glenn, 1975; Lee, Seccombe, & Shehan, 1991). The significant contrast on health status between the two groups accords with the reality which health and happiness could promote each other. Happy people always live longer (Diener & Chan, 2011), and keeping healthy would visibly enhance well-being (Okun, Stock, Haring, & Witter, 1984). Consistent with the results of the previous studies (e.g. Witter, Stock, Okun, & Haring, 1985), we found that religion contributes to the variance in happiness: believers have higher happiness than non-believers.

The use of personal plural pronouns such as "we", "you" (second personal plural, is a Chinese characteristic word category) and "they" which we take to be represents for a focus on relationships are highly correlated with the presence of happiness (Seligman, 2011). Table 3 shows that word categories "achieve" and "work" which have an obviously influence on happiness are tied to achievement, which supports the theory of psychological well-being that the experiences of "self-acceptance" and "personal growth" are sources of happiness (Ryff & Singer, 1996). The lexicon categories suggest a connection with interpersonal relationship such as "social processes" (e.g. mate, talk, greet), "humans" (e.g. adult, baby, boy) and "religion". Previous studies (e.g. Schiffrin, Edelman, Falkenstern, & Stewart, 2010; Valkenburg, Peter, & Schouten, 2006) found that social networks including getting in touch with believers (Lim & Putnam, 2010) could enhance well-being. The other word categories of statistical significance such as "multifunction" (a Chinese characteristic word category, e.g. 有"have", 的"of"), "certain", "space", "preposition" and "nonfluency" which we still have not found the theories or evidences to explain them.

Four behavior features refer to "follow" are shown higher counts in the HH group (see table 4). This indicates that these users interacted with other users more frequently, suggesting a possible growth of social connection. Some studies (e.g. Mitchell, Lebow, Uribe, Grathouse, & Shoger, 2011) revealed that social support obtained from social connection plays an important role in enduring stress and increasing happiness. A large number of status counts have been shown to improve happiness, the reason is that a talkative people equals to more extroverted and extroversion is a trait which could make people easier to be happy. Feature "verified" suggests a rise of the degree of self-disclosure. Christopherson (2007) demonstrated that online anonymity may encourage people to express negative feelings. For this reason, verified users are more restrained on negative expression than anonymous users, which may mislead us that they are happy.

**Prediction.** We observed a better performance of behavior features than linguistic features in the prediction task in Table 5. Figure 1 shows the features which contribute to low happiness. The top three effective factors are all relate to social connection. The prior discussion demonstrates that lack of social support will decreased the ability to resist stress and happiness. From the above we conclude that behavior features, especially social behaviors provides useful information that can be utilized to classify and predict whether an individual may be experiencing a low level of happiness.

# Conclusion

The aim of this study is to see how different levels of happiness exhibit themselves in a social network site and to evaluate predictive proficiency of selected features.

Variance analysis reveals happier people exhibit a "talkative" trait. They post more microblogs and more words about personal plural pronouns, achievement, religion, social contact etc. We know that happiness is linked with income in term of the analysis result of demographics above, but it is interesting that it is "work" and "achieve" that people in happier group talk about, rather than "money". We also found happier people enjoy interacting with other users, and tend to open their identity.

The results of this study shows that we can predict the happiness via a set of data comprise of linguistic and behavior features, but some limitations regarding generalizability are noted. Demographic information suggest that the data distribution of our sample lacks representativeness. Results from a younger, richer, and more urban sample could not be generalized to general population. Language is complex that computer cannot accurately understand it. Base on this reason, our lexicon analysis system is limited to process texts. Nevertheless, this study is proved that it is possible to use social network sites as a tool for measuring and predicting individual happiness. Our work is a small step towards a social media-based happiness perceptron and predictor which can identify happiness as different classes. In further, we hope to improve the ability of our analyzing tool and make more reasonable sampling for estimating happiness more correctly. We also hope this approach can be transformed into the ability of online prediction in the future.